\documentclass[sn-mathphys,Numbered]{sn-jnl}

\usepackage{graphicx}%
\usepackage{multirow}%
\usepackage{amsmath,amssymb,amsfonts}%
\usepackage{amsthm}%
\usepackage{mathrsfs}%
\usepackage[title]{appendix}%
\usepackage{xcolor}%
\usepackage{textcomp}%
\usepackage{manyfoot}%
\usepackage{booktabs}%
\usepackage{algorithm}%
\usepackage{algorithmicx}%
\usepackage{algpseudocode}%
\usepackage{listings}%

\begin{document}

\title[Article Title]{Cactus-like Metamaterial Structures for Electromagnetically Induced Transparency at THz frequencies}


\author*[1,2]{\fnm{Savvas} \sur{Papamakarios}}\email{spapamakarios@iesl.forth.gr}

\author[3]{\fnm{Odysseas} \sur{Tsilipakos}}

\author[1]{\fnm{Ioannis} \sur{Katsantonis}}

\author[1,4]{\fnm{Anastasios D.} \sur{Koulouklidis}}

\author[1]{\fnm{Maria} \sur{Manousidaki}}

\author[1]{\fnm{Gordon} \sur{Zyla}}

\author[1]{\fnm{Christina} \sur{Daskalaki}}

\author[1,5]{\fnm{Stelios} \sur{Tzortzakis}}

\author*[1,5]{\fnm{Maria} \sur{Kafesaki}}\email{kafesaki@iesl.forth.gr}

\author*[1]{\fnm{Maria} \sur{Farsari}}\email{mfarsari@iesl.forth.gr}

\affil*[1]{\orgdiv{Institute of Electronic Structure and Laser}, \orgname{Foundation for Research and Technology-Hellas}, \orgaddress{\city{Heraklion}, \postcode{GR-70013}, \country{Greece}}}

\affil[2]{\orgdiv{Department of Physics}, \orgname{National and Kapodistrian University of Athens}, \orgaddress{\city{Athens}, \postcode{GR-15784}, \country{Greece}}}

\affil[3]{\orgdiv{Theoretical and Physical Chemistry Institute}, \orgname{National Hellenic Research Foundation}, \orgaddress{\city{Athens}, \postcode{GR-11635}, \country{Greece}}}

\affil[4]{\orgdiv{Department of Physics and Regensburg Center for Ultrafast Nanoscopy (RUN)}, \orgname{University of Regensburg}, \orgaddress{\city{Regensburg}, \postcode{93040}, \country{Germany}}}

\affil[5]{\orgdiv{Department of Materials Science and Engineering}, \orgname{University of Crete}, \orgaddress{\city{University of Crete}, \postcode{GR-70013}, \country{Greece}}}


\abstract{THz metamaterials present unique opportunities for next generation technologies and applications, as they can fill the ``THz gap'' originating from the weak response of natural materials in this regime, providing a variety of novel or advanced electromagnetic wave control components and systems. Here, we propose a novel metamaterial design, made of three-dimensional, metallic, "cactus like" meta-atoms, showing electromagnetically induced transparency (EIT) and enhanced refractive index sensing performance at low THz frequencies. Following a detailed theoretical analysis, the structure is  realized experimentally using multi-photon polymerization and electroless silver plating. The experimental characterization results obtained through THz time domain spectroscopy validate the corresponding numerical data, verifying the high potential of the proposed structure in slow light and sensing applications.
}




\maketitle

\section{Introduction}
Electromagnetic metamaterials are artificial media consisting of subwavelength resonant components (meta-atoms) \cite{Lemoult2012} that provide the capability to modulate electromagnetic waves in a wide frequency range even in uncommon ways \cite{Pendry2006}, alternating the electromagnetic properties of conventional materials by given a specific geometry. For many years, metamaterials were used to enable and demonstrate exotic optical properties \cite{aluengheta,Lu2012,Shalaev2007,Yannopapas2009}  in structures composed of simple metallic and/or dielectric materials. More recently, involvement of more complex materials,  e.g., lossy/active media \cite{tsakmakidis_reply,kirby2011evanescent}, magneto-optical materials \cite{almpanis2020spherical} and time-varying media \cite{gardes2007micrometer}, resulted in even more exotic responses. 

Such exotic metamaterial properties directed the scientific interest to the study of a plethora of phenomena, including phenomena that can be observed in quantum systems \cite{Cortes2012,Paspalakis2002158021}. Using metamaterials, quantum phenomena can be exhibited in a macroscopic scale and in a wide range of frequencies \cite{tunablemetamaterial}. Focusing on THz frequencies can provide advanced possibilities for next-generation applications, since the THz band is associated with high potential impact in important application areas,  including sensing, imaging, and communications \cite{Zhu2021,Watts2014,Xu2017,Wang2020334}; moreover  it is highly unexplored as it lies in the border between photonics and electronics.   

Aiming to make a concrete step towards filling the ``gap'' in applications at THz wave manipulation components, we demonstrate here, both theoretically and experimentally, a novel, yet simple, metamaterial design able to exhibit Electromagnetically Induced Transparency \cite{Fleischhauer2005,liu2009plasmonic,liu2010planar} (EIT),  associated also with enhanced sensitivity in THz sensing applications. EIT phenomenon refers to the nonlinear optical property of materials to become transparent in a narrow frequency bandwidth around an absorption line where otherwise they would be opaque. This characteristic performance also results in slowing light and sensing applications through the narrow frequency window that EIT occur. The proposed structure is a two-dimensional array  of pairs of free-standing vertical (three-dimensional) U-shaped Ring Resonators (RRs) \cite{lefthandedmtm}  with an asymmetry  in one of the arms of the U-pair. The resulting geometry resembles a cactus [Figure \ref{fig:Structure}(a)].

Recently, metamaterials exhibiting EIT have attracted considerable interest because of their promising behaviour as low-loss slow-light media. Slow-light media can exhibit very low propagation velocity of the electromagnetic waves \cite{zheng2022dual} or even stop  light entirely \cite{Phillips2001}. 
In EIT,  a medium turns transparent in a narrow frequency window with low absorption and sharp dispersion \cite{wu2015plasmon}, being valuable for slow light and optical storage applications \cite{Fleischhauer2005}. The characteristic features of EIT, simultaneously low absorption and sharp dispersion, can be observed in classical systems with coupled resonators \cite{Singh2009,Ketzaki:2013} and described by a simple model of two coupled harmonic oscillators resulting in dark and bright resonances \cite{Tassin2012} which are excited by breaking the symmetry of the metamaterial, creating states which otherwise the would be forbidden.

Most experimental demonstrations of EIT-like response with metamaterials have been performed with planar meta-atom structures \cite{Yang2014,liu2010planar,Pitchappa2016}. 
Using three-dimensional meta-atoms instead can prove advantageous for certain applications, due to the larger surface area that becomes available for light-matter interaction. For instance, in sensing applications the analyte can occupy a large volume surrounding the 3D meta-atom and interacting with the strong local fields, thus leading to enhanced sensitivity. Note that a sensing application is also naturally suited to the sharp spectral feature of the EIT-like response; not only the sensitivity but also the sensitivity over linewidth ratio (a typical figure of merit in sensing systems) would be enhanced. In addition, 3D meta-atoms allow for more design flexibility in breaking the symmetry with respect to mirror planes of the structure. This can be useful for introducing controllable coupling to already existing dark resonant modes. 
However, meta-atoms with true 3D geometry are challenging from a fabrication standpoint since they cannot be realized with standard lithographic techniques, such as photolithography, e-beam lithography, or nano-imprint lithography. Fortunately, multi-photon polymerization (MPP) offers the unique advantage to realize elaborate 3D structures with practically arbitrary complexity. Focusing in low THz frequencies the sub-wavelength components have to be a few $\mu$m in size. Adding the 3D geometry to the complexity of the structure, MPP enables the fabrication of this kind of structures, while the optical, chemical and mechanical properties of the photosensitive materials can be alternated regarding the requirements of the study. Using MPP provides also the capability for further post-processing techniques such as calcination and metallization process.

In this paper we exploit a 3D metallic ``cactus-like'' structure, as shown in Figure \ref{fig:Structure}(a), to trigger the EIT phenomenon in low THz frequencies ($0.1-10$~THz), and demonstrate its potential for advanced applications in slowing light and environmental refractive index sensing, going one step forward in filling the "THz gap" with new applications about light-matter interaction. The asymmetry takes place in the $yz$ plane, and when the sample is excited with linearly polarized light along the $y$-axis (propagating in the $z$-direction), the quasi-dark appears; its interference with the bright resonance results in a sharp transmission peak within a broad transmission dip. In this study, an extensive explanation of how the EIT phenomenon is triggered in this specific design is elaborated using numerical simulations and a multipole decomposition of the induced conduction current in the structure. Moreover, a simple resistor-inductor-capacitor (RLC) model is provided, able to explain and reproduce not only the response of our structure but also of many different systems where broken symmetry induces sharp EIT-like features.  
The uncommon electromagnetic response of our structure results in a steep change in transmission phase over a narrow frequency window, leading to high values of group delay allowing to delay an incident beam by up to $\sim2200$ optical cycles. In addition, due to the metallic properties and the topology and sensitivity of the resonant modes our structure shows an advanced sensitivity in variations of the refractive index of the environment, being able to reach figures of merit as high as 34. The theoretical design is translated into a real-life structure using MPP as a fabrication tool and subsequently electroless silver plating to get the metallic metamaterial. The results of the theoretical analysis for a specific value of cut in the asymmetry of the ``cactus-like'' design were validated using THz-time domain spectroscopy (TDS), demonstrating the triggering of EIT response in the proposed metamaterial.

\section{Results}
\subsection{Proposed structure and supported resonances}

\begin{figure}[h!]
\centering
\includegraphics[width=13cm]{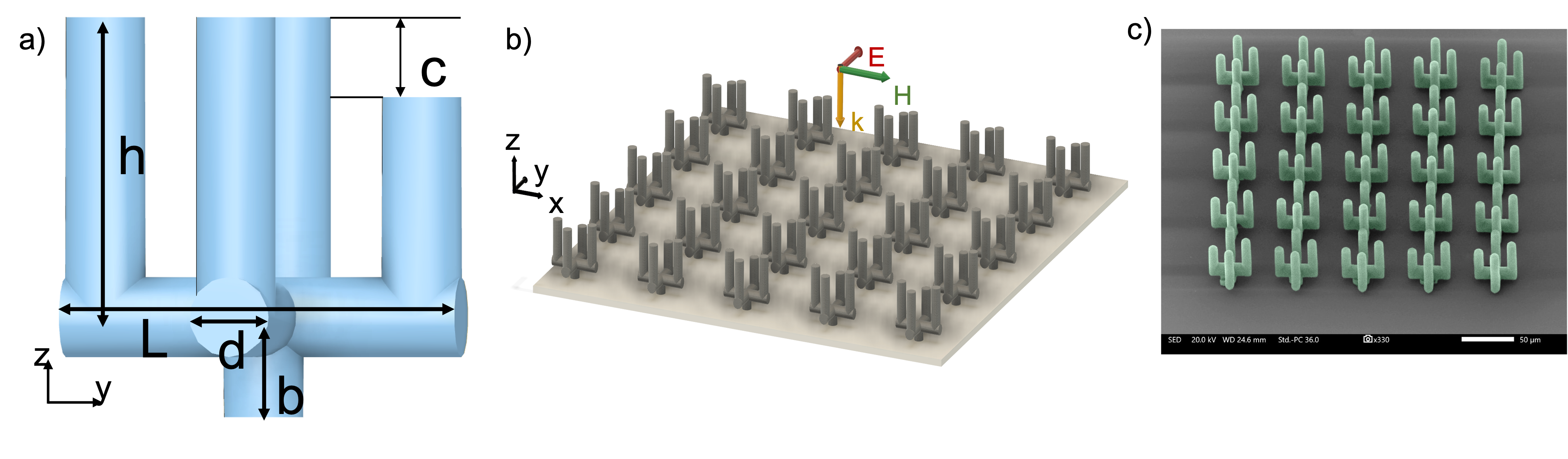}
\caption{Schematic of the proposed metamaterial structure (metasurface). (a) Unit cell with a ``cactus-like'' meta-atom. One arm can be shorter by a value $c$. (b) Metasurface composed of multiple meta-atoms. The incident field is a $y$-polarized normally-incident plane wave.(c) Artificial SEM image of the cactus structures presented for visual purposes. Green color represents the cactus metamaterial and the scattering effect from silver nanoparticles is visualised.}
\label{fig:Structure}
\end{figure}

The proposed metamaterial is illustrated in Figure~\ref{fig:Structure}(b). It consists of a  periodic arrangement (square lattice in the $xy$ plane) of 3D meta-atoms with a ``cactus-like'' geometry [Figure~\ref{fig:Structure}(a)]. Each meta-atom is formed by two metallic, vertically-positioned U-shaped split-ring resonators (SRRs) placed in the $xz$ and $yz$ planes, respectively. In addition, one of the two SRRs (the one in the $yz$ plane) can be asymmetric, when one vertical arm is cut and shorter by a value $c$ from the other one. This specific microstructure offers the ability to control all the important parameters of symmetric and asymmetric features and it covers a large volume area of the unit cell which is crucial for sensing applications since the light-matter interaction is increased. The asymmetric features gives full control of the EIT phenomenon ,and, subsequently to the group delay and sensitivity of the metamaterial. The metallic behavior of the structure offers enhanced interaction with THz radiation and higher sensitivity in refractive index changes. These two characteristics of geometry and material's properties are critical parameters to excite EIT in low THz frequencies. The metal that was used is Silver which provides high conductivity and it is suitable for low THz frequencies regarding the light-matter interaction.

\begin{figure}[]
\centering
\includegraphics[width=12cm]{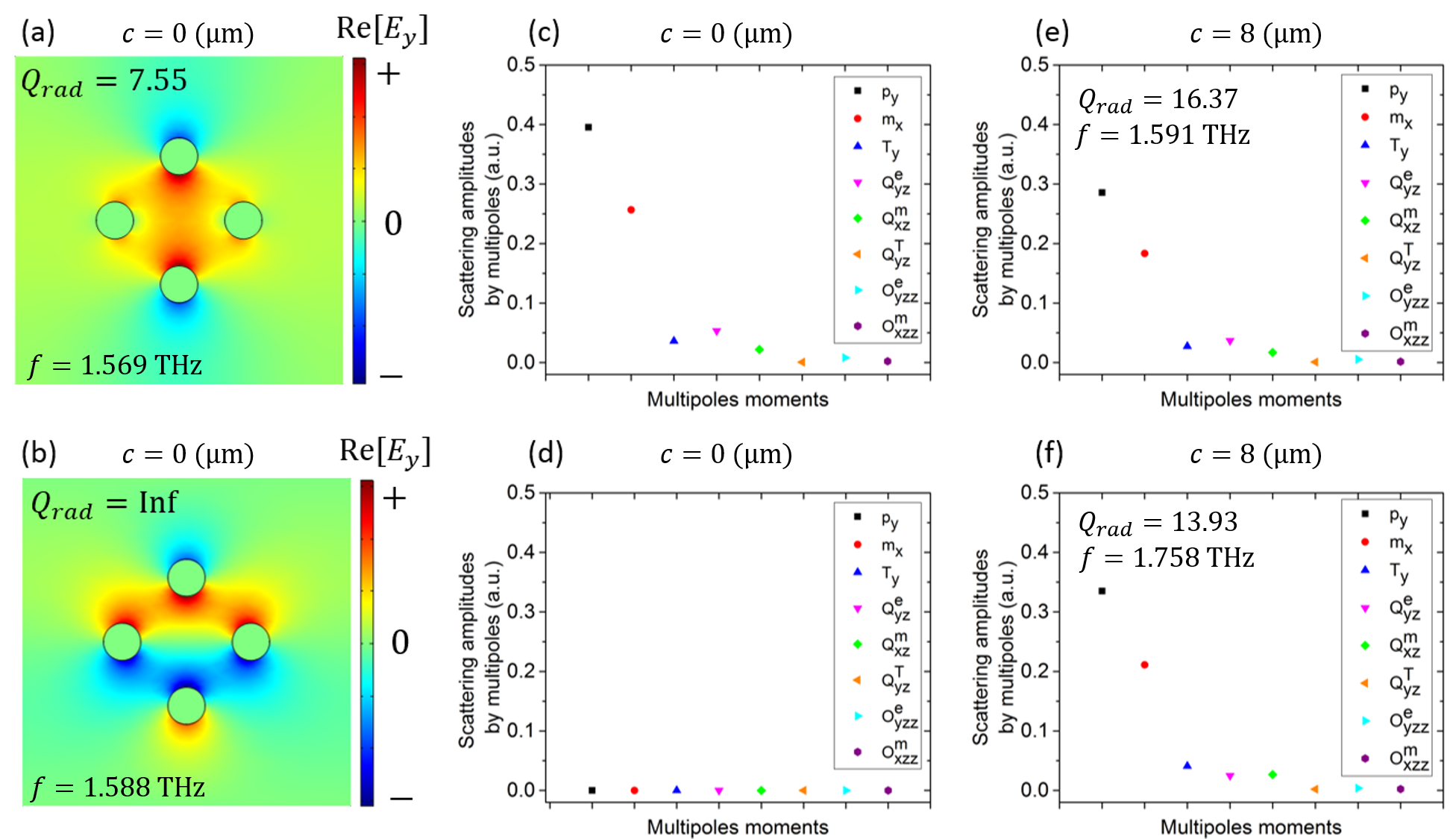}
\caption{Bright and dark eigenmodes supported by the ``cactus'' meta-atom. (a),(b) Electric field distribution ($E_y$ component) at a $xy$ plane bisecting the meta-atom for the (a) bright and (b) dark mode in the symmetric $c=0$ case. The resonant frequencies and $Q_\mathrm{rad}$ values are also included.  
(c),(d)~Multipole decomposition of the conduction current distribution in the meta-atom, indicating the contribution of each multipole moment to the $y-$polarized scattered field. In the dark mode [panel (d)], all contributions are near-zero. 
(e),(f) Same as panels (c) and (d) when the meta-atom becomes asymmetric with $c=8~\mu$m. The second mode becomes quasi-dark and now exhibits considerable $p_y$ and $m_x$ components.}
\label{fig:Eigenmodes}
\end{figure}

Starting our investigation from the symmetric meta-atom (no cut, $c=0$) and solve an eigenvalue problem of the periodic unit cell to determine the resonant modes of the metasurface.  For the simulations, it was considered a 50~nm  silver coating layer with conductivity $\sigma_{Ag}=6.3\times10^6$~S/m. The dimensions of the meta-atom are $h= 40~\mu$m, $L= 40~\mu$m, $d= 9~\mu$m and the unit cell size is $80~\mu$m. The substrate and support leg are momentarily omitted to simplify the structure and focus on the underlying physics.
Concentrating on the two modes, which are depicted in Figure~\ref{fig:Eigenmodes}(a),(b). Besides the electric field distribution ($E_y$ component), the resonant frequencies and radiation quality factors, $Q_\mathrm{rad}$, are also included. It can be seen that the modes lie close in frequency but they exhibit quite different $Q_\mathrm{rad}$ values. 
The mode in Figure 2(a) is bright, i.e, it possess a finite and relatively low $Q_\mathrm{rad}$ and can be directly accessed with a $y$-polarized normally-incident plane wave. This is corroborated by the even parity of the $E_y$ component with respect to the $xz$ and $yz$ planes. On the other hand, the mode in Figure 2(b) is dark; it possesses a practically infinite $Q_\mathrm{rad}$ and cannot couple to a normally-incident $y$-polarized plane wave. This is corroborated by the odd parity exhibited by the $E_y$ component with respect to the $xz$ plane.   

Next, we investigate further the electromagnetic character of each mode by performing a multipole decomposition \cite{Savinov2014}. The far-field contribution of each multipole to a $y-$polarized scattered field \cite{Tsilipakos:2018} is presented in Figure~\ref{fig:Eigenmodes}(c) and (d) for the two modes under study. In the bright-mode case [Figure~\ref{fig:Eigenmodes}(c)], we see strong electric ($p_y$) and magnetic ($m_x$) dipole contributions, indicating that both the $E_y$ and $H_x$ components of a $y$-polarized plane wave can excite the eigenmode. In addition, we find non-negligible contributions by the electric quadrupole $Q_{yz}^e$ and toroidal dipole $T_y$, since the finite dimensions and elaborate geometry of the meta-atom lead to a more complex near-field structure of the resonance. In sharp contrast, for the dark mode [Figure~\ref{fig:Eigenmodes}(d)] we find near-zero contributions to a $y-$polarized scattered field for all multipole moments. This is yet another confirmation of the dark nature of the mode. 

We now examine the role of asymmetry by introducing a cut of $c=8~\mu$m in one of the arms [see Figure~\ref{fig:Structure}(a)]. The frequencies of the two modes shift, as anticipated, and the new resonant frequencies are $f=1.591$~THz and $f=1.758$~THz, respectively. More importantly, due to the broken symmetry, the dark mode now becomes quasi-dark \cite{Tsilipakos:2021apl}. This strategy has received renowned interest recently in the context of quasi bound states in the continuum (qBIC) \cite{Zografopoulos:2023}. Specifically, the mode profile is not strictly antisymmetric with respect to the $xz$ plane anymore, the radiative quality factor becomes finite ($Q_\mathrm{rad}\sim 13.93$ in this case), and excitation via a normally-incident $y-$polarized plane wave is allowed. Note that the symmetry with respect to the $yz$ plane has not been disturbed. As discussed in the following, the excitation of the quasi-dark mode introduces a sharp transmission peak within a broad transmission dip, leading to a spectral response reminiscent of electromagnetically-induced transparency. This significant  change in the radiative characteristics is imprinted in the multipole expansion as well [Figure~\ref{fig:Eigenmodes}(f)]; the $p_y$ and $m_x$ contributions are not negligible anymore and can mediate coupling with an incoming $y$-polarized plane wave (e.g., incident wave or local $E_y$ and $H_x$ fields produced by the originally bright mode). Slight changes in the multipole composition are also seen for the bright mode [Figure~\ref{fig:Eigenmodes}(e)]. In the supplementary material (SM), we track the evolution of the multipole contributions with varying $c$, providing also the corresponding frequencies and Q-factors. In addition, we devise a simple RLC circuit model able to capture the  geometrical features of the structure, considering also the excitation configuration, and provide the resulting dark and bright modes and their dependence on the structure characteristics. This model, among others, demonstrates how two dissimilar coupled resonators supporting ``bright'' resonances that can be excited externally by the same ``source'' result  to EIT-like response.  

\subsection{EIT response and sensing performance}

\begin{figure}[h!]
\centering
\includegraphics[width=12.8cm]{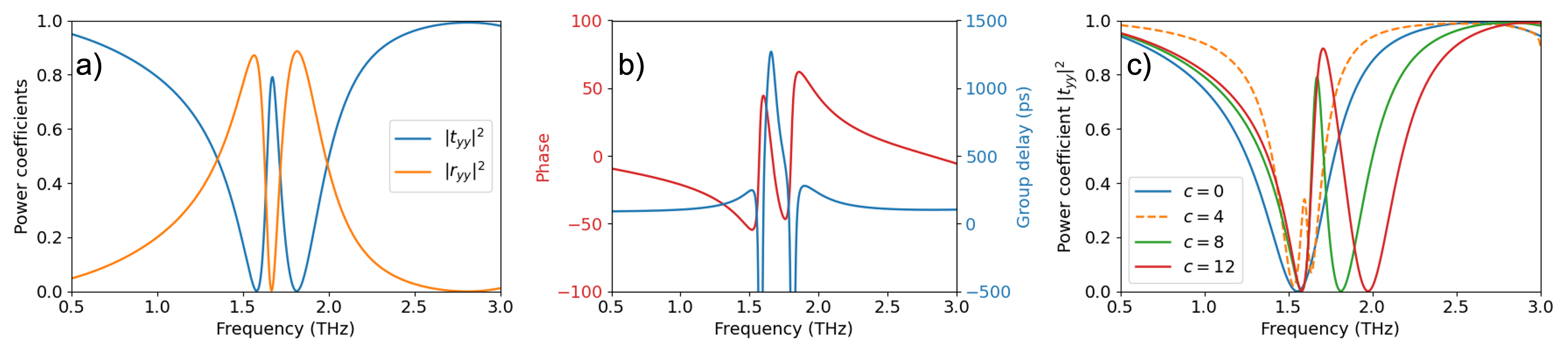}
\caption{Simulated results of the electromagnetic response for the proposed structure for $y$-polarized light. (a) Reflection (red curve) and transmission (blue curve) power coefficients ($R_{yy},T_{yy}$)  for asymmetry parameter $c=8$ $\mu$m. (b) Transmission phase (red curve) and group delay (blue curve) for  $c=8$ $\mu$m. (c) Transmission (power) coefficient for different values of $c$; the EIT feature arises when $c\neq 0$.}
\label{fig:T_phase_delay}
\end{figure}

We next calculate plane-wave scattering coefficients for the metasurface under study, by means of full-wave simulations based on the finite element method. We assume that a $y$-polarized plane wave impinges on the metasurface  at normal incidence.
Transmission and reflection coefficients for the $c=8~\mu$m case are depicted in Figure \ref{fig:T_phase_delay}(a).  

As discussed in Figure~\ref{fig:Eigenmodes}(f), in the asymmetric structure ($c\neq 0$), the quasi-dark mode can be excited. It interferes with the bright mode and, as a result, a sharp transmission peak appears within the broad transmission dip; the quasi-dark mode resonant frequency emerges at $f\sim1.758$~THz. The interference between the two modes is further discussed in the SM by performing a multipole expansion on the induced conduction current and identifying that both the bright and quasi-dark modes are characterized predominantly by $p_y$ and $m_x$ contributions (cf. Figure~\ref{fig:Eigenmodes}). Note that in the symmetric structure ($c=0$), the antisymmetric mode is completely dark, and a single conventional transmission dip appears associated with the bright resonance (see SM). In the same way, when the incident wave is $x-$polarized, the quasi-dark mode cannot be excited, and a single transmission dip appears in the scattering coefficients (see SM).

The response in Figure \ref{fig:T_phase_delay}(a) is reminiscent of the quantum phenomenon of EIT and is, thus, frequently termed ``photonic analogue of EIT'' \cite{Ketzaki:2013,Yu2018}. This analogy is further corroborated by looking at the transmission phase in Figure \ref{fig:T_phase_delay}(b). Indeed, there is a central region of steep phase delay with negative slope. This region can be used for delaying light \cite{He2015}, as can be seen by the group delay which is calculated as $\tau_g=-{d\phi(\omega)}/{d\omega}$ and reaches a value of 1.3~ns (approximately 2200 times the carrier cycle at 1.7~THz). The characteristics of the EIT feature can be readily controlled by varying the degree of asymmetry (parameter $c$). This is depicted in Figure \ref{fig:T_phase_delay}(c). As $c$
increases, the quasi-dark mode becomes brighter and the EIT peak becomes broader. At the same time, the maximum group delay decreases but the bandwidth that can be used for pulse delaying purposes increases.  

\begin{figure}[]
\centering
\includegraphics[width=14cm]{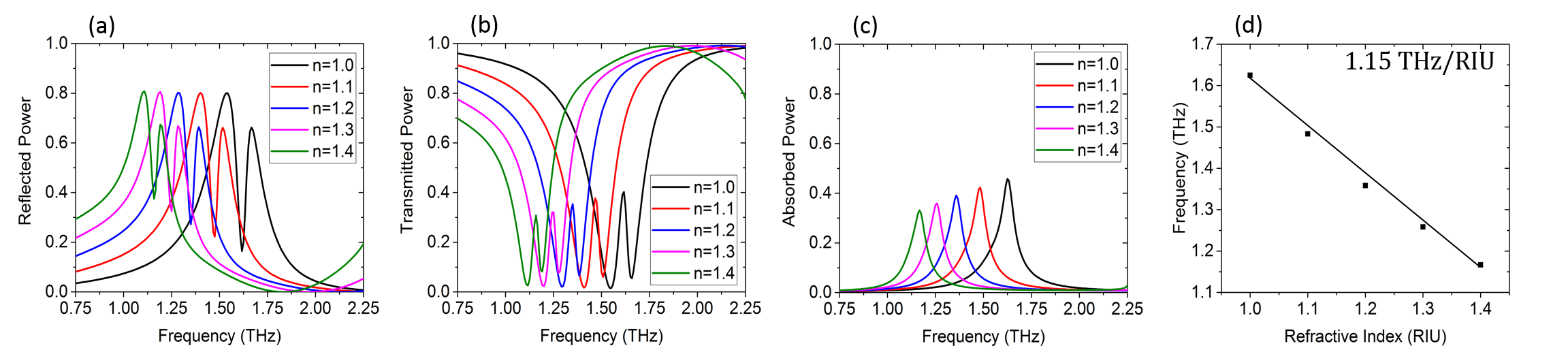}
\caption{Sensitivity of our proposed metamaterial design for different values of refractive index. Panels (a) reflected power, (b) transmitted power, c) absorbed power and d) resonance shifts for refractive index variation from $n=1$ to $n=1.4$. }
\label{fig:sensing}
\end{figure}

Next, we investigate the merits of the proposed structure as a refractive index (RI) sensor. As mentioned in the introduction, the 3D meta-atom allows for increased surface area for light-matter interaction \cite{xomalis2022enhanced}. We choose a small degree of asymmetry ($c=4~\mu$m), so that the EIT peak is sharp. We assume that the analyte exactly covers the meta-atoms and conduct simulations for a varying RI value for the analyte in the range $n=1-1.4$ (relevant for biological media). The reflection,  transmission, and absorption coefficients are depicted in Figure \ref{fig:sensing}(a), (b), and (c), respectively. The sharp EIT feature shifts linearly as the RI of the analyte changes, and its position in either transmission or reflection can be readily traced. This linear shift is shown in Figure~\ref{fig:sensing}(d). From the slope, we can deduce a sensitivity value of $S=\Delta f/\Delta n=1.15$ THz/RIU \cite{sherry2006localized}. This high sensitivity can also be attributed to the strong local fields and the 3D nature of the metallic ``cactus'' meta-atom \cite{xomalis2022enhanced}. In addition, to better assess the sensing performance, we introduce the typical figure of merit ($\mathrm{FOM}=S/FWHM$), which takes into account the linewidth of the EIT feature. The full-width at half-maximum (FWHM) is determined via the transmission curve. For the $c=4~\mu$m case studied in Figure \ref{fig:sensing}, the FOM can be calculated to 34 which is higher than recent works has shown in electromagnetic metamaterials operating in THz frequencies in combination with the enhanced sensitivity \cite{li2024high,liu2010planar,pan2019terahertz,Zhu2021}.

\subsection{Fabrication of 3D metamaterials to induce EIT at THz frequencies}
The experimental validation of the proposed metamaterial is realized using MPP. Given its nonlinear characteristics, MPP enables true and maskless 3D printing with sub-micron resolution, a capability highly beneficial for various tasks across interdisciplinary research areas \cite{zyla_generation_2017,flamourakis2020laser,fischer2013three,bertoncini20203d,ottomaniello2023highly,gonzalez2023micro}. These characteristics make MPP an ideal tool for the fabrication of metallic "cactus-like" resonators with both symmetric and asymmetric designs and experimentally verifying their optical properties. Although MPP mainly facilitates the processing of dielectric structures, the use of a photoresist containing a precursor with moieties capable of binding metals enables the selective metallization of MPP-processed structures in a post-processing step. This metallization process, achieved via a highly selective chemical approach known as silver electroless plating (SEP), which deposits silver nanoparticles exclusively where these moieties exist, i.e., on the surfaces of the structures, transforming them into conductive ones, as detailed described in Refs. \cite{sakellari20173d, Katsantonis2023}.

The MPP fabrication of the structure was done using  a hybrid photoresist with Zr-based inorganic component \cite{Ovsianikov2008} (see SM). The photoresist shows dielectric properties and a post-metallization process was required to obtain conductive structures suitable for the low THz regime. After a carefully designed SEP protocol (see SM) based on previous work that has been done for various applications, we achieved silver-nanoparticle-coating of final thickness, overcoming the skin depth of silver in low THz frequencies.  

\begin{figure}[h!]
\centering
\includegraphics[width=13cm]{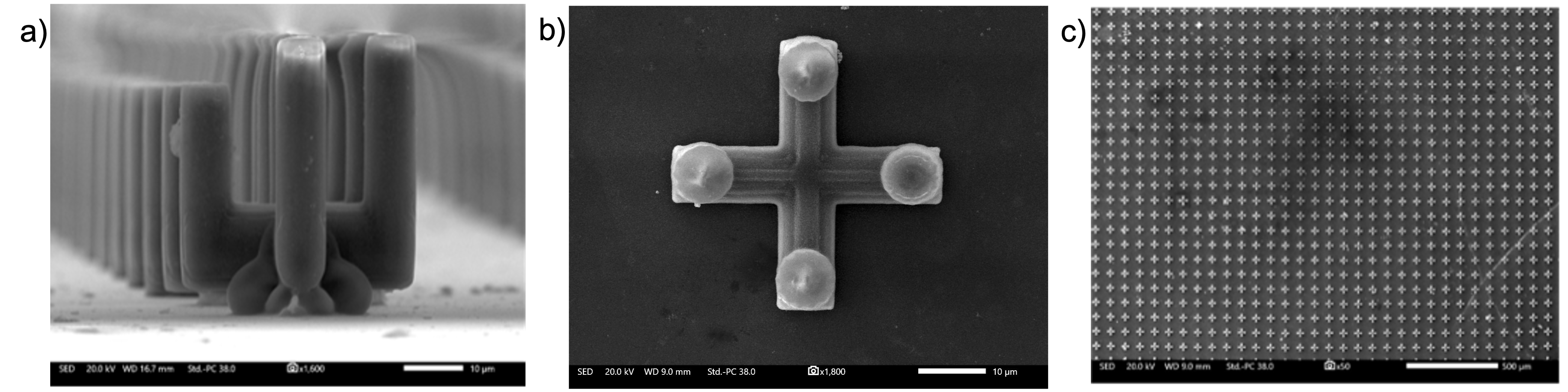}
\caption{SEM images of the fabricated metamaterial in (a) side view (unit cell), (b) top view (unit cell), (b) top view (array)}
\label{SEM_images}
\end{figure}

Regarding the dimensions of the fabricated structures, the unit cell size was 80$\times$80 $\mu$m$^2$ in the $xy$ plane). The entire processed area comprises approximately 1700 unit cells and occupies $\sim3.4\times3.4$ mm\textsuperscript{2} in order to overcome the THz beam diameter in the characterization process which is 2~mm. The manufacturing of this area required approximately 7 hours. The resulting 3D U-shaped resonators are depicted in Figure \ref{SEM_images}, which shows SEM images with different magnifications and orientations of the structure. All SEM images distinctly illustrate the high stability and uniformity of the structures [Figure \ref{SEM_images}(a),(c)], key characteristics for achieving reliable functionality in a metamaterial. The deformed bottom base in single meta-atoms, which arose due to the reflective properties of silicon against 780~nm laser radiation enhancing the polymerization of the material close to the substrate, are minor structural imperfections insignificant for the working principle of the proposed metamaterial. This is because all meta-atoms exhibit the same defect, and this defect does not affect the electromagnetic structure response. Note that the EIT mechanism relies on the asymmetry of one U-shape resonator, where one arm is shorter than the other one, as shown in Figure \ref{SEM_images}(a); this feature was perfectly processed.  

\begin{figure}[h!]
\centering
\includegraphics[width=13cm]{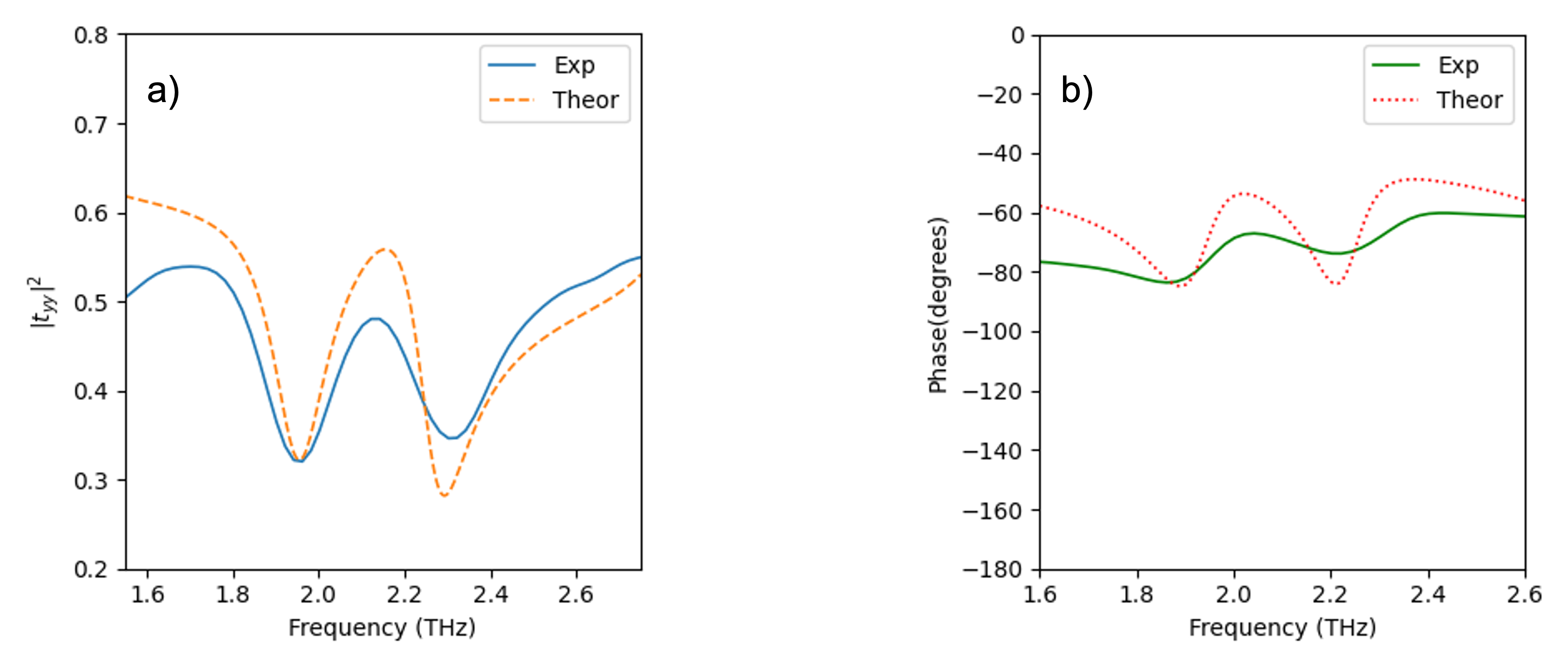}
\caption{(a) Transmission (power) coefficient for linearly polarized light along y axis ($T_{yy}=|t_{yy}|^2$). (b) Corresponding transmission phase. Blue curves show the experimental data, red curves the corresponding theoretical ones.}
\label{fig:exp_results_T_phase}
\end{figure}

In order to detect the response of the fabricated sample in low THz frequencies (1-5 THz), Time Domain Spectroscopy (TDS) \cite{naftaly2007terahertz} is used. The electric signal is obtained using photoconductive antennas (PCAs) \cite{lepeshov2017enhancement} and the detector is placed behind the silicon substrate. By rotating the sample we were able to detect linearly polarized wave along the $y$ axis and measure the transmission coefficient $T_{yy}$. This transmission, along with the corresponding theoretical result, is shown in Figure \ref{fig:exp_results_T_phase}.

For the theoretical result, in order to get the optimum fitting to the experiment,  now the calculations were done with the dimensions and the parameters of the fabricated structures. From SEM images in Figure \ref{SEM_images}, the dimensions were measured to be $d=8.9$ $\mu m$, $L=40$ $\mu m$, $\alpha=79.8$ $\mu m$ the size of the unit cell (lattice constant), $h=35.2$ $\mu m$, and $c=8.1$ $\mu m$. The silver nanoparticles that coat the structure were measured to be approximately 150 nm in size, using SEM images with high magnification (see SM). This difference in dimensions of the structure as well as the thinner silver coating of the polymer creates a shift in the resonances of the EIT to higher frequencies as it was predicted also from the parametric study of different pillar heights (see SM). In addition, the final simulations take into account the final sample properties. The silicon substrate with relative permittivity of $\epsilon_s=11.2$ and loss tangent $\tan\delta=0.02$ is included in the calculations. The conductivity of the silver coating is set to be $\sigma_{exp}=5.75\times10^5$~S/m, as found by conductivity measurements 
(see SM); this value is lower than the one that has been measured previously in the literature \cite{Vasilantonakis2012}. The lower conductivity is attributed to the multiple repeats of the metallization process in order to reach the desired thickness. This generates a multilayer system of silver nanoparticles, which decrease the mobility of the carriers on the surface \cite{Brown1994} (similar to the single-layer and multi-layer graphene system \cite{Nagashio2009}). 
In addition, the surface roughness (which may be a few nanometers) can effectively leads to lower conductivity through scattering effects \cite{chen2007modeling}. 

Lower conductivity results in reduced transmission of the THz radiation through the sample as theoretical parametric simulations for different silver conductivity values have shown (SM). Thus, the intensity of the transmission was expected to be relatively low. However, the EIT phenomenon was still observable and not affected by the conductivity of the metallic structure, since it is triggered by the broken symmetry imported in the system and not significantly by the electric properties of it.

From the TDS characterization we obtained an electric signal in time and using Fourier transformation we get the amplitude/phase in the frequency domain. Dividing the detected signal of the bare silicon substrate with the manufactured one, we obtain the normalised transmission spectra for the proposed metamaterial (SM). As mentioned already, the experimental results for transmission, as well as the theoretical fitting from the simulations are presented in Figure \ref{fig:exp_results_T_phase}(a). From the experimental curve (blue curve) we observe that despite the increased losses, the characteristic EIT spectral feature is clearly visible. This is verified by the theoretical calculations (orange curve) incorporating the experimental conditions, which show good agreement with the experimental data. In addition, the EIT nature of the response is also corroborated by the transmission phase, depicted in Figure \ref{fig:exp_results_T_phase}(b). 

The behavior of the EIT response from the experiment comes in alignment with the theoretical predictions from the aspect of where the resonances are expected to be with the exact dimensions of the fabricated structure and the experimental conditions included in the simulations. 
In the experimental part we observe the dark resonance at $f=1.95$~THz and the bright resonance at the transmission peak of EIT at $f=2.13$~ THz. Nevertheless, the narrow EIT window (the two dark resonances are separated by a transparency window of $f=0.43$~ THz) shows promising results for delaying light applications. In addition, transmission phase was measured again using TDS and the final results for transmission phase which triggers also the delaying light applications as it was described in subsection 2.2 are exhibited in Figure \ref{fig:exp_results_T_phase}(b). 

The enhanced losses observed in the characterization of the structure can be explained due to the metallization process, as it was described above, substrate’s resistivity, non-negligible amount of dirt and silver nanoparticles on the substrate which are responsible for the absorption of a small part of the THz radiation before it reaches the detector. In addition, the silver nanoparticles are not formed completely homogeneous onto the polymerized material.

\section{Discussion}
In this paper, we have demonstrated the design and fabrication of a metallic metamaterial exhibiting electromagnetically induced transparency (EIT) associated with enhanced refractive index sensing performance in low THz frequencies. Our design resembles a cactus, featuring two vertical metallic U-shaped rings arranged perpendicular to each other. One of the rings has a broken symmetry, crucial for inducing a coupling between a dark and a bright resonance, leading to the EIT feature.  The structure was realized experimentally using MPP and selective electroless silver plating. Its electromagnetic response was experimentally measured through THz time domain spectroscopy and has been verified against numerical simulations taking into account actual dimensions and realistic material parameters. In the theoretically optimized metal-coated structure we achieved a transmission amplitude of 80\% at the EIT peak, along with a group delay of 1.3 ns ($\sim2200$ carrier cycles). The investigation of the refractive index sensing performance of the structure yielded  a quite high figure of merit (FOM $\approx$ 34). This advanced sensing potential originates from the  sharp EIT feature combined with the 3D meta-atom geometry associated with large surface area for the structure-analyte interaction. 
The proposed structure and  research underscores the potential of 3D-meta-atom configurations for advanced applications in slowing light and  environmental sensing, offering insights also into the design and fabrication challenges associated with 3D meta-atom-based metamaterials.

\section*{Materials and methods}

A suitable photoresist for processing via MPP and subsequent post-processing via SEP is a metal-binding photopolymer which 2-(dimethylamino) ethyl methacrylate is added at a concentration of 30\% v/v relative to the base ZPO monomers \cite{Ovsianikov2008, Farsari2010}. For its use in this study, the modified photoresist was synthesized in-house (see SM for details) and then drop-casted onto high-resistivity silicon substrates (thickness: 540~$\mu$m, resistivity: 100-1000 $\Omega\cdot$cm) that exhibit semi-transparent optical properties in the THz region of interest. Prior to deposition, the silicon substrates were silanized to enhance adhesion between the processed structures and the silicon surface. For better attachment of structures on the substrate, a monolayer of $3-(Trimethoxysilyl)propyl methacrylate$ (MAPTMS) was formed on the surface of the substrate following a silanization process. The substrates were immersed in a solution of Ammonium hydroxide ($NH_4OH$) and Hydrogen peroxide ($H_2O_2$) at a volume ratio $3:1$ and heated at $75^o$ for 15 minutes in order to clean the surface. Subsequently, they were immersed in distilled water and dried. Finally, the sinalization of the substrates is completed by immersing them in a solution of Toluene and MAPTMS at 0.5\% $v/v$ and let them overnight. Next, the substrates are cleaned using ethanol or acetone and stored in ethanol in cool and dark environment. 40 $\mu L$ of the photosensitive material are drop-casted using a pipette. The drop-casted material was kept under low vacuum conditions at room temperature for 2 days, ensuring the complete evaporation of any residual solvents. After the fabrication process, the sample was immersed for 45 minutes in $4methyl-2pentanone$ and the rinsed in $isopropanol$ for another 30 minutes. Here, in order to achieve the selectivity of SEP, it is important to remove all the excess non-polymerized material from the sample and the substrate in order to avoid the deposition of silver nanoparticles onto them.
    
The sample was processed using the optical setup depicted in the SM. The setup for the fabrication of the proposed metamaterial structure primarily includes the irradiation source for the MPP process which is a Femtosecond Fiber Laser (FemtoFiber pro NIR, Toptica Photonics AG) emitting at a central wavelength of 780~nm, with pulse duration 150~fs, average output power 500~mW and repetition rate 80~MHz. Also, a 2D Galvo scanner system (Scanlabs HurryscanII 10) that it is consisting of galvanometric mirrors scans the laser beam on $xy$ plane during the fabrication process, an acousto-optical shutter, and a high-resolution xyz axis system. A microscope objective lens with $40x$ magnification (Zeiss, Plan Apochromat)  with numerical aperture 0.95 was employed to focus the laser beam onto the photoresist. The fabrication of single 3D U-shaped resonators was carried out in a layer-by-layer approach, following a bottom-to-top strategy, using a slicing and hatching distance of 1 $\mu m$ and 500 nm accordingly. More specifically, the process started with the fabrication of the taller pillars, followed by the arms and finishing at the bottom base of the structure, using a galvo scanning velocity of 3 mm/s and an average laser power of 100 mW, as measured by a digital power meter positioned in front of the last mirror before the galvo scanner system. 

The metallization process that was used is a modification of the one established in \cite{Vasilantonakis2012} and further demonstrated in \cite{Aristov2016,Katsantonis2023}. SEP is a straightforward chemical procedure that does not require any apply of electrical potential and provides silver nanoparticles coating in a highly selective way only on the surface of the polymerized material. In THz frequencies the skin depth of silver is approximately 80 nm \cite{kang2018terahertz}, and thus the protocol was modified to fulfill this condition. From SEM images depicted in the Supplementary Material, the formed coating on the surface of the polymerized material was roughly 150 nm. Since the coating is not completely homogeneously placed the thickness can only be estimated using the mean value of the size of silver nanoparticles that are formed on the surface. Energy Dispersive X-ray Spectroscopy (EDX) experiments were also done in order to observe the presence of silver on the fabricated structure.

\section*{acknowledgements}
Authors acknowledge funding by European Union through projects MSCA CHARTIST (GA No. 101007896),  FABulous (HORIZON-CL4-2022-TWIN-TRANSITION- 01-02, GA:101091644). This work was supported by the Marie Skłodowska-Curie Actions, under grant agreement No. 101059253, as part of the European Union's Horizon Europe research and innovation programme. The research project was co-funded by the Stavros Niarchos Foundation (SNF) and the Hellenic Foundation for Research and Innovation (H.F.R.I.) under the 5th Call of “Science and Society” Action – “Always Strive for Excellence – Theodore Papazoglou” (Project Number: 9578.). Authors acknowledge also, Ms Aleka Manousaki for the excellent SEM support.

\section*{conflict of interest}
The authors declare that there is no conflict of interest.

\begin{appendices}




\end{appendices}


\bibliography{references}

\end{document}